\documentstyle[12pt]{article}

\newcommand{\beq}{\begin{equation}}
\newcommand{\eeq}{\end{equation}}
\newcommand{\beqa}{\begin{eqnarray}}
\newcommand{\eeqa}{\end{eqnarray}}
\newcommand{\dfrac}{\displaystyle \frac}

\begin{document}


\begin{center}
{\large\bf CHIRAL PREDICTION FOR THE
{\boldmath $\pi N$} S--WAVE

\vspace{0.2cm}

 SCATTERING LENGTH
{\boldmath   $a^-$} TO ORDER {\boldmath ${\cal O}(M_\pi^4)$}}
\end{center}

\vspace{.5in}

\begin{center}
V. Bernard$^a$\footnote{also at Laboratoire de Physique Th\'eorique,
Institut de Physique, 3-5 rue de l'Universit\'e, F-67084 Strasbourg
Cedex, France.}\footnote{electronic address: bernard@crnhp4.in2p3.fr},
N. Kaiser$^b$\footnote{electronic address: nkaiser@physik.tu-muenchen.de},
Ulf-G. Mei{\ss}ner$^c$\footnote{electronic address:
meissner@pythia.itkp.uni-bonn.de}\\

\bigskip

\bigskip


\it
$^a$Centre de Recherches Nucl\'eaires, Physique Th\'eorique\\
BP 28, F-67037 Strasbourg Cedex 2, France\\
\vspace{0.2cm}
$^b$Technische Universit\"at M\"unchen, Physik Department T39\\
James-Franck-Stra\ss e, D-85747 Garching, Germany\\
\vspace{0.2cm}
$^c$Universit\"at Bonn, Institut f{\"u}r Theoretische Kernphysik\\
Nussallee 14-16, D-53115 Bonn, Germany\\
\end{center}

\vspace{1in}

\thispagestyle{empty}

\begin{abstract}
We evaluate the S-wave pion--nucleon scattering length $a^-$ in the
framework of heavy baryon chiral perturbation theory
up--to--and--including terms of order
$M_\pi^4$. We show that the order $M_\pi^4$ piece of
the isovector amplitude at threshold, $T^-_{\rm thr}$, vanishes
exactly. We predict for the isovector scattering length,
$0.088 \, M_{\pi^+}^{-1} \le a^- \le 0.096 \,
M_{\pi^+}^{-1}$.
\end{abstract}

\vfill

\today



\newpage

\noindent Chiral symmetry severely constrains the interactions of the
strongly interacting particles at low energies, first formulated in
terms of current algebra (CA). One of the most splendid successes of
CA was Weinberg's prediction for the S--wave pion--nucleon scattering
lengths \cite{wein},
\beq
a^+_{\rm CA} = 0 \, \, , \quad a^-_{\rm CA} = \dfrac{M_\pi}{8\pi
  F_\pi^2} \dfrac{1}{1+ M_\pi/m_p } = 0.079 \, M_\pi^{-1} \, \, ,
\label{CA}
\eeq
with $M_\pi = 139.57$ MeV the charged pion mass, $m_p = 938.27$ MeV the
proton mass, $F_\pi = 92.5$ MeV the pion decay constant and the
superscripts $+/-$ refer to the isoscalar and isovector $\pi N$
amplitude, respectively.
The Karlsruhe--Helsinki phase shift analysis of $\pi N$ scattering
\cite{karl} leads to
$a^- = 0.092 \pm 0.002 \, M_\pi^{-1}$ and $a^+= -0.008 \pm 0.004 \,
M_\pi^{-1}$, impressively close to the CA prediction, eq.(\ref{CA}).
The Karlsruhe--Helsinki analysis is based on dispersion relations thus
requiring a solid data base as input.
However, over the last few years there has been some controversy about
the low--energy $\pi N$ data which has not yet been
settled. Consequently, the uncertainties in $a^{\pm}$ are presumably
larger and even the sign of $a^+$ could be positive. A more direct way
to get a handle at these zero momentum (i.e. threshold) quantities is the
measurement of the strong interaction shift ($\epsilon_{1S}$) and the
decay width ($\Gamma_{1S}$) in pionic atoms.
The PSI-ETH group has
recently presented first results of their impressive measurements in
pionic deuterium  and pionic hydrogen \cite{bad} \cite{pisd}
\cite{sigg}  (for earlier work in the field see e.g. \cite{forster},
\cite{bovet} \cite{beer}). The S--wave
scattering length are related to $\epsilon_{1S}$ and $\Gamma_{1S}$ via
the Deser--type formulae \cite{deser},
\beq
\epsilon_{1S} = - C (a^+ +a^-) (1+\delta_\epsilon) \, , \quad
\Gamma_{1S} = 4\, q \, C (1 + 1/P) \bigl[ \, a^-
\,(1+\delta_\Gamma ) \bigr]^2 \, \, ,
\label{des}
\eeq
with $q$ the momentum of the $\pi^0$, $P$ the Panofsky ratio and the
constant $C= 4E_{1S}/r_B$ is proportional to
the point-Coulomb binding energy, $E_{1S}$, of
the system in the $1S$ state divided by the pion Bohr
radius, $r_B$. $\delta_\Gamma$ and $\delta_\epsilon$ describe electromagnetic
corrections to the scattering lengths, see \cite{siggl}. The
consequent analysis of the data
leads to \cite{sigg}
$a^- = 0.096 \pm 0.007 \, M_\pi^{-1}$ and $a^+= -0.0077 \pm
0.0071 \, M_\pi^{-1}$. If one combines the pionic hydrogen shift measurement
with the one from the pionic
deuterium, one has
$a^- = 0.086 \pm 0.002 \, M_\pi^{-1}$ and $a^+= 0.002 \pm 0.001 \,
 M_\pi^{-1}$. The
largest uncertainty comes from the width measurement of pionic
hydrogen. Both determinations are consistent within one standard
deviation. We conclude that $a^-$ is larger than the CA value and that
$a^+$ is consistent with zero.

Heavy baryon chiral perturbation \cite{jm} allows to systematically calculate
the corrections to the CA predictions, eq.(\ref{CA}), for the
S--wave scattering lengths. This topic was already addressed
in ref.\cite{bkma}. In this note, we want to further sharpen the
prediction for the isovector scattering length. In particular, we will
calculate all contributions which are of order $M_\pi^4$ and
furthermore give a realistic estimate of the uncertainties for the one
loop result (which was not done in \cite{bkma}). The starting point is
the effective pion--nucleon Lagrangian,
\beq
{\cal L}_{\pi N} = {\cal L}_{\pi N}^{(1)} + {\cal L}_{\pi N}^{(2)} +
{\cal L}_{\pi N}^{(3)} + {\cal L}_{\pi N}^{(4)}
\label{leff}
\eeq
where the superscript $(i)$ refers to the number of derivatives or
quark mass insertions. The structure of ${\cal L}_{\pi N}$ is
discussed in detail in the review \cite{bkmr} and a pedagogical
introduction can be found in \cite{prag}.
The S-wave scattering lengths can be calculated from the
 on--shell forward $\pi N$
scattering amplitude $T^{ba} = T^+(\omega) \delta^{ba} + i \epsilon^{bac}\tau^c
T^-(\omega)$  with $\omega = p\cdot q/m_p$ the pion lab energy. Evaluation at
threshold $\omega = M_\pi$ gives them as $a^\pm = T^\pm_{\rm thr} / (4 \pi ( 1
+M_\pi/m_p)) $.
The one-loop result for $T^-_{\rm thr}$ to order $M_\pi^4$ reads
\beqa
T^-_{\rm thr} = {M_\pi \over 2 F_\pi^2}
+ {g_{\pi N}^2 M_\pi^3 \over 8 m_p^4}
+ {M_\pi^3\over 16 \pi^2 F_\pi^4}
\biggl( 1 - 2 \ln {M_\pi \over \lambda} \biggr)
 +B^r(\lambda)\,M_\pi^3
 + T^{-,4}_{\rm thr} + {\cal O}(M_\pi^5)\, , \label{tmin}
\eeqa
where the chiral corrections at order $M_\pi^3$ come from the
expansion of the nucleon pseudovector Born term, the loop corrections and the
counterterms. $T^{-,4}_{\rm thr}$ denotes the order $M_\pi^4$
contribution to the isovector forward $\pi N$ amplitude at
threshold to be discussed below.
We note that the contribution from the Born term is
just 1$\%$ of $a^-_{\rm CA}$. To determine the
counterterm contribution, one needs to pin down
the value of $B^r (\lambda)$ which is a particular combination of
some low--energy constants from
${\cal L}_{\pi N}^{(3)}$ \cite{bkma} \cite{bkmr}.
In fact, these low-energy constants are not
known and have been estimated in \cite{bkma} via resonance
exchange,
\beq
B^r (\lambda) = B^{\rm Res} =
{g_{\pi N}^2 \over 2 m_p^2 m_\Delta^2 } \biggl(Z - \dfrac{1}{2}
\biggr)^2 + {R g_{\pi N}^2  \over 8 m_p^2(m_p + m_{N^*})^2 } \, \, .
\label{reso}
\eeq
 In eq.(\ref{reso}), we have explicited the
corresponding $\Delta(1232)$ and the $N^*(1440)$ contributions. The
range of the various resonance parameters like $R$ and $Z$ is
discussed in \cite{bkma}.
This procedure induces a spurious dependence on the scale of
dimensional renormalization ($\lambda$) which enters through the
divergences from the one loop contribution. In \cite{bkma}, $\lambda$
was fixed at the mass of the $\Delta (1232)$ since this resonance is
most important in the estimation of the contact term $B^r (\lambda)$.
However, in the framework of resonance saturation this  scale should be
varied between  $M_\eta = 550$ MeV and $m_{N^*} = 1.44$ GeV.
Would we be able to fix  $B^r (\lambda)$ from some data, the
prediction for $a^-$ would, of course, be $\lambda$--independent.

We now turn to the calculation of the order $M_\pi^4$
corrections. First, contact terms from ${\cal L}_{\pi N}^{(4)}$ can not
contribute to $T^-_{\rm thr}$ at ${\cal O}(M_\pi^4)$
 because of the crossing property of
the isovector forward $\pi N$ amplitude,
\beq
T^- (\omega) = - T^- (- \omega) \quad , \label{cro}
\eeq
with $\omega$ the pion lab energy. Thus we are left with one--loop graphs
with exactly one insertion from ${\cal L}_{\pi N}^{(2)}$. Many of
these diagrams cancel with their crossed partners
due to the crossing property, eq.(\ref{cro}), and specific properties
of the loop functions entering,
as exemplified in Fig.1. In particular, the terms with one insertion
proportional to $c_{1,2,3,4}$ from ${\cal L}_{\pi N}^{(2)}$ vanish
because
\beq
J_0^{\pi N} (\omega) + J_0^{\pi N} (- \omega)=
-\dfrac{1}{4\pi}\sqrt{M_\pi^2 -\omega^2 } = 0 \, \, ,
\eeq
at threshold, $\omega = M_\pi$ (the pertinent loop functions are given
in appendix B of \cite{bkmr}).
However, this does not hold for the thirteen diagrams shown in
Fig.2. These remaining graphs can be grouped into five classes as
indicated in Fig.2. Straightforward calculation gives
\beq
T^{-,4}_{\rm thr} = \dfrac{g_A^2 \,  M_\pi^4}{128 \pi \,  m_p \, F_\pi^4} \,
\biggl(9 + 8 - 2 + 5 - 20 \biggr)
= 0 \, , \label{zero}
\eeq
where we have exhibited the contributions from the various classes separately.
This result is reminsicent of the
non--renormalization of the isovector charge of the
nucleon. Consequently, the next chiral corrections appear at order
$M_\pi^5$, from two--loop graphs, one loop graphs with one insertion
from ${\cal L}_{\pi N}^{(3)}$, counterterms and so on. They are
suppressed by four powers of $M_\pi / (4\pi F_\pi)$ compared to the
leading order term, eq.(\ref{CA}) and are thus expected to be small,
even though the exact numerical coefficients can be of order ten (as
indicated from the ${\cal O}(M_\pi^3)$ correction to the CA result).

Varying the various resonance parameters which enter the estimate of
$B^{\rm Res}$ and the scale $\lambda$ within their bounds, we find as
a conservative estimate for the one--loop prediction of $a^-$,
\beq
0.088 \, M_{\pi^+}^{-1} \le a^- \le 0.096 \,  M_{\pi^+}^{-1} \, \, ,
\label{amc}
\eeq
which is consistent with the various empirical values discussed
before and $10 \ldots 20\%$ larger than the CA prediction \cite{wein}.
As already stressed in \cite{bkma}, it is the chiral loop
correction at order $M_\pi^3$ which closes the gap between the lowest
order (CA) prediction and the empirical value. An indication of the
size of the next corrections can be obtained by writing the one--loop
result as $a^- = a^-_{\rm CA} (1 + \delta_1) \simeq a^-_{\rm CA}
 \exp(\delta_1)$. The next correction follows to be $\delta_1^2 / 2$ which
 is of the order of $1 \ldots 2\%$ of $a^-_{\rm CA}$. We stress that
 this should only be considered indicative. The uncertainty in the prediction
 eq.(\ref{amc}) could be further decreased if one could determine $B^r
 (\lambda )$ from some other process.

To summarize, we have sharpened the chiral perturbation theory
prediction for the isovector $\pi N$ S--wave scattering length $a^-$
by showing that the term of order $M_\pi^4$ in $T^-_{\rm thr}$ is
identical  to zero. Varying all
uncertainties entering the one--loop expression for $a^-$ \cite{bkma},
we arrived at the band, eq.(\ref{amc}), which is consistent with the
existing data. Concerning the isoscalar scattering length $a^+$, no
firm prediction can be given at present as discussed in \cite{bkma}
since there are large cancellations which make the result for $a^+$
very sensitive to some not very accurately known counterterms.

\vspace{1cm}

\noindent{\Large {\bf Acknowledgements}}

\medskip

\noindent We are grateful to J\"urg Leisi and Daniel Sigg for triggering our
interest and communication of results before publication.

\vspace{1cm}

\noindent{\Large {\bf Figure Captions}}

\bigskip

\begin{enumerate}

\item[Fig.1] Cancellation of diagrams at order $M_\pi^4$
  due to the crossing property,
  eq.(\ref{cro}). Solid and dashed lines denote nucleons and pions,
  respectively. The heavy dot denotes an insertion from
  ${\cal L}_{\pi N}^{(2)}$.

\medskip

\item[Fig.2] Classes of non--vanishing diagrams at order
  $M_\pi^4$. For notations, see Fig.1.

\end{enumerate}


\begin{thebibliography}{99}

\bibitem{wein}S. Weinberg, Phys. Rev. Lett. {\bf 17}, 616 (1966)

\bibitem{karl}R. Koch, Nucl. Phys. {\bf A448}, 707 (1986)

\bibitem{bad}A. Badertscher et al., Nucl. Instr. $\&$ Methods {\bf
    A335}, 470 (1993)

\bibitem{pisd}D. Chatellard et al., ``Determination of the S-wave
  scattering length in pionic deuterium with a high resolution crystal
  spectrometer'', Jan. 1995, submitted to Phys. Rev. Lett.

\bibitem{sigg}D. Sigg, Ph.D. Thesis ETH No. 11049, Z\"urich, 1995
(unpublished)

\bibitem{forster} A. Forster et al.,  Phys. Rev {\bf C28}, 2374 (1983)

\bibitem{bovet} D. Bovet et al, Phys. Lett.
{\bf B153}, 231 (1985)

\bibitem{beer} W. Beer et al, Phys. Lett.
{\bf B261}, 16 (1991)

\bibitem{deser}S. Deser, M.L. Goldberger, K. Baumann and W. Thirring,
Phys. Rev {\bf 96}, 774 (1954)

\bibitem{siggl}D. Sigg et al., ``Electromagnetic Corrections to the
  S-Wave Scattering Lengths in Pionic Hydrogen'', subm. for
  publication

\bibitem{jm}E. Jenkins and A.V. Manohar, Phys. Lett. {\bf B255}, 558
  (1991)

\bibitem{bkma} V. Bernard, N. Kaiser and Ulf-G. Mei\ss ner, Phys. Lett.
{\bf B309}, 421 (1993)

\bibitem{bkmr} V. Bernard, N. Kaiser and Ulf-G.~Mei\ss ner, ``Chiral
Dynamics in Nucleons and Nuclei'', preprint CRN 95/3 and TK 95 1,
Int. J.Mod. Phys. E (1995), in print

\bibitem{prag}Ulf--G. Mei{\ss}ner, Czech. J. Phys. {\bf 45}, 153
  (1995)


\end{thebibliography}
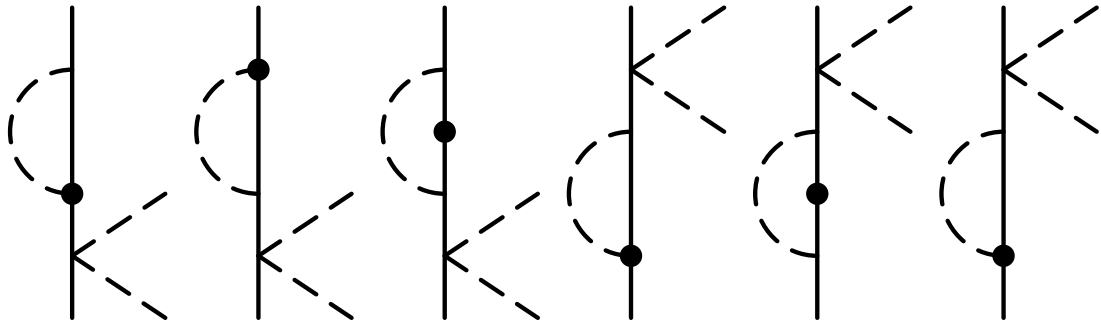
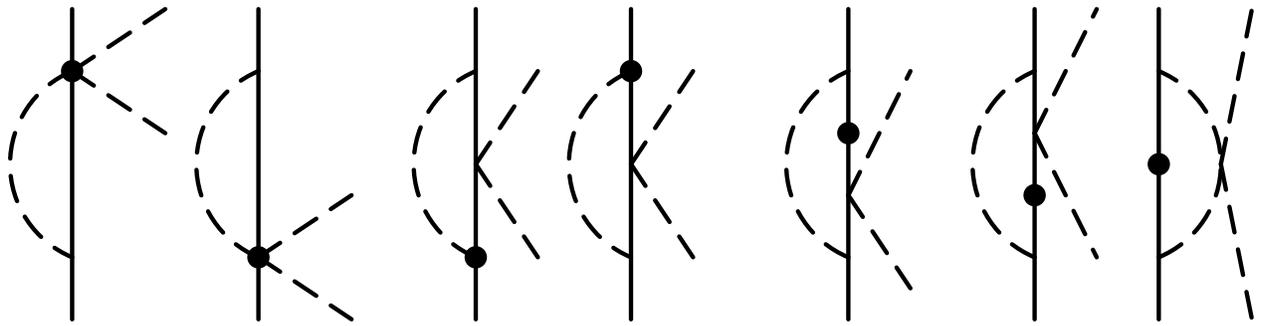
\end{document}